\documentclass[aps,prb,reprint,superscriptaddress,amsmath,amssymb,showkeys,a4paper]{revtex4-1}

\usepackage[utf8]{inputenc}
\usepackage{setspace}
\usepackage{graphicx}
\usepackage{dcolumn}
\usepackage{bm}
\usepackage[T1]{fontenc}  
\usepackage{lmodern}   
\usepackage{subfigure}
\usepackage{siunitx} 
\usepackage{epsf}
\usepackage{xcolor}
\usepackage{soul}

\begin{document}

\title{Approaching the Trap-Free Limit in Organic Single Crystal Field-Effect Transistors}
	\author{B. Bl\"ulle}
	\email{bbluelle@phys.ethz.ch}
	\author{R. H\"ausermann}
  	\author{B. Batlogg}
\affiliation{Laboratory for Solid State Physics, ETH Zurich, 8093
Zurich, Switzerland}

\date{\today}

\begin{abstract}
We present measurements of rubrene single crystal field-effect transistors with textbooklike transfer characteristics, as one would expect for intrinsically trap-free semiconductor devices. Particularly, the high purity of the crystals and the defect-free interface to the gate dielectric are reflected in an unprecedentedly low subthreshold swing of $65$ ${\rm mV / decade}$, remarkably close to the fundamental limit of $58.5\,{\rm mV / decade}$. From these measurements we quantify the residual density of traps by a detailed analysis of the subthreshold regime, including a full numerical simulation. An exceedingly low trap density of $D_{bulk} = 1 \times 10^{13}~{\rm cm^{-3}eV^{-1}}$ at an energy of $\sim0.62~{\rm eV}$ is found. This result corresponds to one trap per eV in $10^8$ rubrene molecules. The equivalent density of traps located at the interface ($D_{it} = 3 \times 10^{9}~{\rm cm^{-2}eV^{-1}}$) is as low as in the best crystalline ${\rm SiO_2/Si}$ field-effect transistors. These results highlight the benefit of having van der Waals bonded semiconducting crystals without electronically active states due to broken bonds at the surface.
	
\end{abstract}

\maketitle 

\section{Introduction}

Charge transport in semiconductors is strongly influenced by the presence of traps, energetically located in the band gap between the two transport levels. The quantification of the density of these trap states (trap DOS) is a crucial step towards understanding the electrical properties of the materials. There are various ways to determine the trap DOS experimentally, which range from photoelectron spectroscopy \cite{Bussolotti2013} over electron spin resonance spectroscopy \cite{Mishchenko2012} to direct measurements of the transport properties of a semiconductor device \cite{Powell1992}. A field effect-transistor (FET) is a well suited device to study the trap DOS, as the spectral distribution of charge traps can be studied through changing the Fermi level by applying a bias to the gate contact. This method has become a powerful tool to study material properties in the field of organic semiconductors \cite{Kalb2010}. 

In the past three decades, organic field-effect-transistors (OFET) have come a long way from the first organic thin-film-transistor\cite{Ebisawa1983}  (OTFT) to single crystal OFETs \cite{Horowitz1996} with mobilites surpassing 40~cm$^2$/Vs. \cite{Takeya2007} The first generation of OFETs do not have any clearly distinguishable subthreshold regime, where an exponential dependence of the drain current on the applied gate voltage is expected. Through advancements in thin-film deposition methods it is possible to build OTFTs with a well-pronounced subthreshold regime, comparable to inorganic amorphous transistors.\cite{Lin1997, Sirringhaus1999} High-purity single crystals of organic molecules \cite{Laudise1998} led to OFETs with a steeper subthreshold slope. \cite{Podzorov2003a} To unleash the full  performance of these single crystals, a compatible gate dielectric is required,  which does not introduce additional charge traps in the semiconductor. More recently, it has been shown that either an air-gap structure \cite{Podzorov2004,Xie2013} or a fluorinated polymer results in high performance OFETs \cite{Kalb2007,Willa2013} with a high mobility and a steep turn-on, enabling fast switching speed and low power consumption. To further improve the turn-on characteristics, there has been a focus on thin and high-$\kappa$ dielectric layers to increase the gate capacitance, leading to a lower subthreshold swing.\cite{Zschieschang2012, Ma2008, Liu2009, Walser2009a, Park2012,Ono2013} 

In this study, we focus on the subthreshold regime from a microscopic perspective, especially its relation to the trap DOS. Single-crystal OFET measurements with extremely low subthreshold swing are presented, and the theoretical description of the subthreshold current to extract the trap DOS is summarized. Furthermore, a method is derived to estimate the Fermi energy at the turn-on voltage and thus the depth of these traps. The range of validity of this analysis is assessed using a full numerical simulation. With either method, we consistently find an extremely low density of deep trap states for rubrene, as low as in crystalline inorganic semiconductors.

\section{Rubrene single crystal FETs}
\begin{center}
\begin{figure}
\includegraphics[width=0.7\columnwidth]{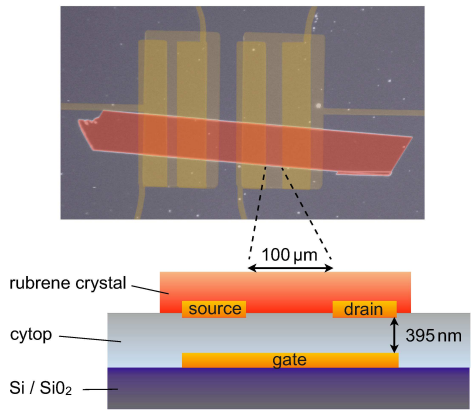}
\caption{ \label{fig:FET_device_picture} Colored photograph and schematic cross-section of the single crystal OFET. The channel length and width are $100\,{\rm \mu m}$ and $270\,{\rm \mu m}$, respectively. The spin-coated Cytop film is $395\, {\rm nm}$ thick, resulting in a gate capacitance of $4.71\, {\rm nF/cm^2}$ ($\varepsilon_{cytop} = 2.1\times\varepsilon_0$). The thickness of the rubrene crystal is $2.3\, {\rm \mu m}$.}
\end{figure}
\end{center}

As a result of extensive studies on the quality of organic FETs, we find that crystalline semiconductors in combination with a highly hydrophobic insulating surface leads to best results.\cite{Kalb2007} To explore the intrinsic limits of molecular semiconductors, we build a series of rubrene single crystal FETs in a bottom-gate/bottom-contact configuration (Fig. \ref{fig:FET_device_picture}) with the amorphous fluoropolymer Cytop as gate dielectric. Cytop CTL-809M (Asahi glass, Bellex Int.) is mixed 1:1 with the solvent CT-Solv-180, spin coated on a pre-cleaned $\rm Si/SiO_2$ substrate and cured at $80^\circ\,\rm{C}$ for $30\,\rm{min}$ and at $120^\circ\,\rm{C}$ for $1\,\rm{h}$. The substrate handling, the spin spin coating and Cytop curing are done in ambient air.  Evaporated $\rm Cr/Au$ and $\rm Au$ layers are used as structured gate and source/drain electrodes, respectively. The rubrene single crystals are grown by physical vapor transport \cite{Laudise1998} from 98\% pure source material (Sigma-Aldrich) without any additional purifying steps and they are attached to the prefabricated substrates by flip crystal technique\cite{Takeya2003a,deBoer2003} in ambient air under a white-light microscope. Thereafter, we transfer the samples to a helium-filled glove box to perform the electrical measurements. The OFETs are measured at room temperature ($T$ = 295~K), by using a HP 4155 A semiconductor parameter analyzer operated with instrument control. \cite{Pernstich2012}

All devices show very similar electrical characteristics: Mobilities extracted from the saturation and the linear regime range from $10$ to $15\,{\rm cm^2/Vs}$ and on-off ratios above $10^7$ at $V_g = -10 \, {\rm V}$ are reached. A common feature is the extremely steep turn-on behavior with a subthreshold swing $S$ in the range of $65$ to $80\,{\rm mV / decade}$. None of the devices show any hysteresis, i.e. no gate bias stress, implying that there is no long-term charge trapping in the OFET\cite{Bobbert2012,Hausermann2011}.  The gate leakage current is below the noise level of the measurement setup at  $200\, \rm fA$. In the linear regime, the drain currents scale linearly with the gate voltage, which is a sign of a negligible small charge injection barrier at the contacts.
Here, we discuss the transistor with the lowest subthreshold swing (Fig. \ref{fig:FET_device_picture}). The rubrene crystal was laminated in ambient air and  measured in helium atmosphere at room temperature ($T = 295 \, {\rm K}$), using a HP 4155A semiconductor parameter analyzer operated with instrument control (iC). \cite{Pernstich2012}

\begin{figure}
\includegraphics[width=1\columnwidth]{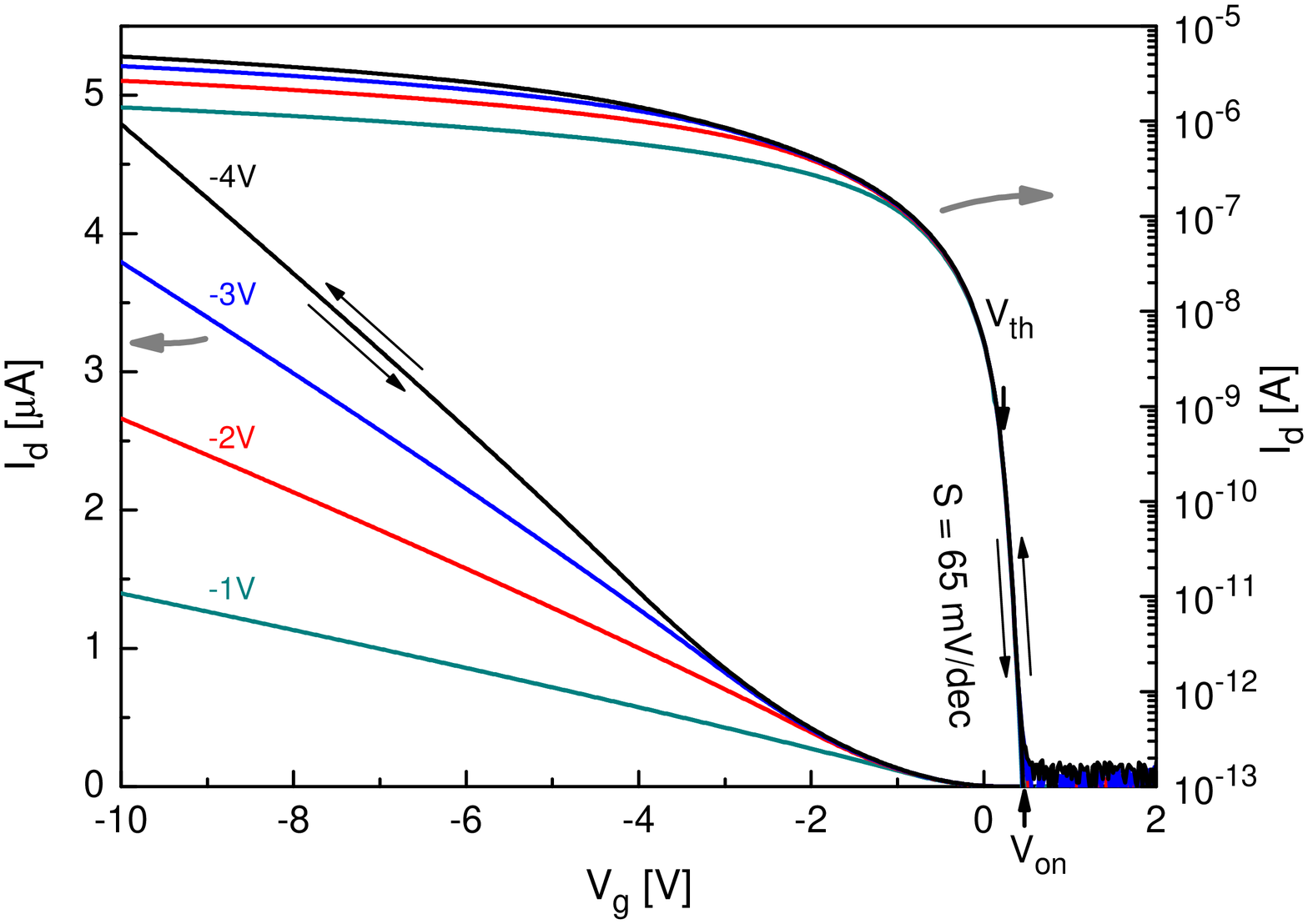}
\caption{ \label{fig:FET_transfer} Measured transfer curve of the rubrene single-crystal FET using a Cytop layer as gate dielectric. The device exhibits no hysteresis and its on-off ratio is larger than $10^7$. The extracted mobility is $\mu_{\rm sat} = 13.9\,{\rm cm^2/Vs}$.  In the subthreshold regime (above $V_{th} = 0.23\,{\rm V}$), the exponential dependence of the currents on the gate voltage corresponds to a subthreshold swing of $S = 65\,{\rm mV/decade}$.}
\end{figure}

\begin{figure}
\includegraphics[trim = 0mm 0mm -24mm 0mm,clip,width=1\columnwidth]{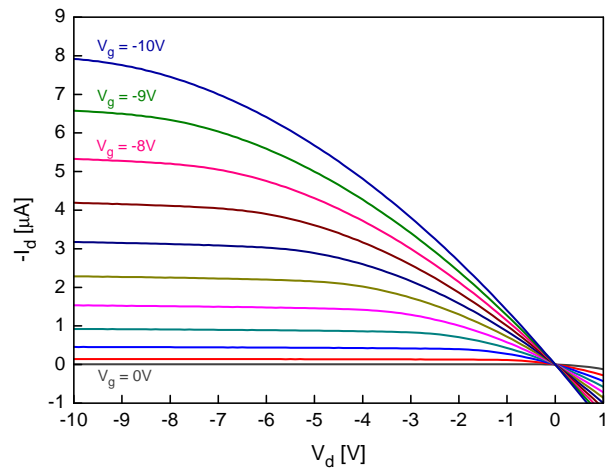}
\caption{ \label{fig:FET_output} Output curves of the rubrene single crystal FET at various values of the gate voltage. The linear zero-crossing (no S-shape) is an indication for a negligibly low charge injection barrier at the contacts.}
\end{figure}

The transfer-curves for various drain voltages and the output characteristics are shown in Fig. \ref{fig:FET_transfer} and \ref{fig:FET_output}. The 2-point field-effect mobilities derived from the linear and the saturation region, $\mu_{\rm lin} = 13.0\,{\rm cm^2/Vs}$ and $\mu_{\rm sat} = 13.9\,{\rm cm^2/Vs}$, are remarkably high.

\begin{figure}
\includegraphics[width=1\columnwidth]{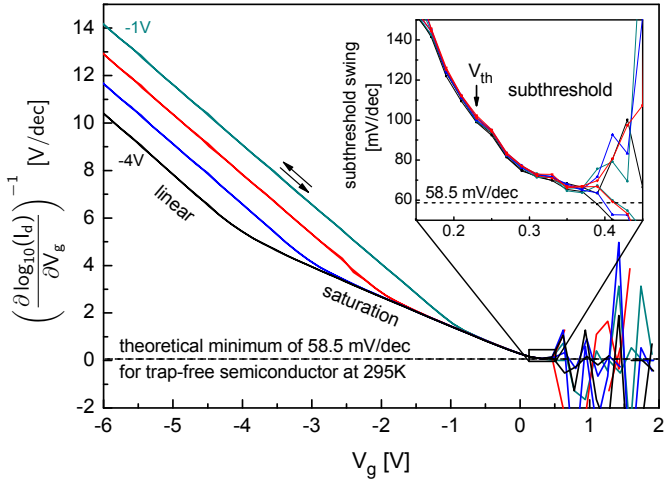}
\caption{Inverse logarithmic slope of the measured transfer curves from Fig. \ref{fig:FET_transfer}. In the main panel, linear and saturation regimes manifest themselves in the known way: $I_d \propto V_g$ and $I_d \propto V_g^2$, respectively. A magnification of the subthreshold region is shown in the inset. With increasing gate voltage, the inverse slope approaches a subthreshold swing value of $65\,{\rm mV / decade}$, very close to the theoretical limit of $58.5\,{\rm mV / decade}$.\label{fig:FET_subswing}}
\end{figure}

In the following we focus on the subthreshold regime which is defined as the region between the turn-on and the threshold voltage, in this device given as $V_{on} =  0.47\,{\rm V}$ and $V_{th} = 0.23\,{\rm V}$ (from saturation regime). In this regime, the drain current increases exponentially with the gate voltage, which is defined as the subthreshold \textit{slope} (in units decade/V) or its inverse, the subthreshold \textit{swing} $S$ (V/decade). The latter is best extracted from a plot of the inverse logarithmic slope of the drain currents versus gate voltage (Fig. \ref{fig:FET_subswing}). In this plot the extremely steep exponential turn-on behavior becomes apparent: In the subthreshold region (Fig. \ref{fig:FET_subswing} inset), the curves truncate at a minimal value of $S = 65 \pm 2\,{\rm mV / decade}$, which is the subthreshold swing. The same value for $S$ is obtained from both sweep directions (no hysteresis) and lies remarkably close to the theoretical limit of $S = 58.5\,{\rm mV / decade}$ at $295\,{\rm K}$.

\section{Subthreshold swing and trap density}

In the subthreshold region, where the gate voltage is below the threshold voltage, the formation of a pinch-off zone with a very low charge carrier density near the drain contact leads to a suppression of the drift current. The large gradient of the charge concentration between the source and drain contact regions, however, gives rise to a diffusion current which is independent of the drain voltage, as long as the drain voltage is larger than a few $k_B T / q$.\cite{Sze1981physics,colinge2002physics}
The subthreshold current is proportional to the carrier concentration, which varies exponentially with the gate voltage. Thus,
\begin{equation}
	I_d \propto exp\left( \frac{qV_g}{n^* k_B T} \right)\,,
\end{equation}
where the so-called subthreshold slope depends on the thermal energy $k_B T/q$ and the ideality parameter $n^*$.\cite{RollandDiss}
The subthreshold swing $S$ is defined as the inverse of the subthreshold slope\cite{Sze1981physics} and corresponds to the gate voltage needed to increase the drain current by a factor of $10$:
\begin{equation}
	S = \frac{ k_B T \ln(10)}{q}\,n^* \,.
	\label{eq:S}
\end{equation}
The ideality parameter $n^*$ is associated with the density of charge traps far away from the transport level, located either at the semiconductor-insulator interface or in the bulk of the semiconductor. Because of their relatively large trapping energy, these states are called \textit{deep} traps. The parameter $n^*$ can be written as \cite{Sze1981physics}
\begin{equation}
	n^* = 1+C_{sc}/C_{i}\,,
	\label{eq:CscCi}
\end{equation}
where $C_i$ is the capacitance of the gate dielectric per unit area. The quality of the semiconducting material expresses itself in an effective capacitance $C_{sc}$, since the filling of trap states while the Fermi energy is pushed towards the transport level is equivalent to the charging of a capacitor.\cite{Kim2013} $C_{sc}$ is distinct from the geometric capacitance of the semiconductor, $C_{sc}^{geom} = \varepsilon_{sc}/t_{sc}$. While $C_{sc}^{geom}$ does not express itself in the DC transfer characteristics, $C_{sc}$ affects the subthreshold swing. Rolland et. al.\cite{Rolland1993} have shown it to be directly dependent on the density of deep trap states in the bulk and at the interface:
\begin{equation}
	\label{eq:Csc}
	C_{sc} = q \sqrt{\varepsilon_{sc} D_{bulk}} + q^2 D_{it} \,,
\end{equation}
where $D_{bulk}$ is the bulk trap density per volume and energy, $D_{it}$ denotes the interface trap density per unit area and energy and $\varepsilon_{sc}$ is the permittivity of the semiconductor material. 

For an ideal transistor without any traps, $C_{sc}$ is zero and the parameter $n^*$ equals $1$. Thus, there is a theoretical limit for the subthreshold swing, given by $S_{ideal} = k_B T \ln(10)/q$ which is $58.5\,{\rm mV/decade}$ at $295\,{\rm K}$.

For a real semiconductor, however, the subthreshold swing is larger than this minimum and the difference between the theoretical minimum and the measured subthreshold swing is a measure of the imperfection of the transistor interface and the semiconductor material.

From the subthreshold swing alone, it is not \emph{a priori} possible to distinguish between trapping at the interface and trapping in the depletion zone of the bulk, since both contribute to an effective capacitance $C_{sc}$, corresponding to a trap concentration per \emph{unit area}. However, we can estimate the maximum density of interface traps contributing to the measured subthreshold swing by setting $D_{bulk}$ to zero:
\begin{equation}
	\label{eqn:Nintmax}
	D_{it}^{max} = \frac{C_i}{q^2} \left(\frac{q S}{k_B T \ln(10)} - 1 \right)\,.
\end{equation}
If at least part of the trap states are located in the bulk, a reasonable assumption for the channel thickness is necessary to convert the areal density into a volume density. In general, this thickness is given by the Debye length $\lambda = \sqrt{\varepsilon_{sc}/ q^2 D_{bulk}}$, which was also used in the derivation of eq. (\ref{eq:Csc}).\cite{RollandDiss} Again, by setting $D_{it}=0$ we obtain the maximum contribution of bulk traps:
\begin{equation}
	\label{eqn:Nbulkmax}
	D_{bulk}^{max} = \frac{C_i^2}{\varepsilon_{sc} q^2} \left(\frac{q S}{k_B T \ln(10)} - 1 \right)^2 \,.
\end{equation}
Care must be taken when the above relations are applied to nearly trap-free semiconductors or to very thin semiconducting layers as in evaporated or solution-processed FETs. For such devices, $\lambda$ can be of the same order as the semiconductor thickness $t_{sc}$ and the charge carriers accumulate almost uniformly throughout the semiconductor. In this situation, the charge transport in the subthreshold region is essentially a volume phenomenon rather than an accumulation of charge carriers within the first few monolayers of the semiconductor\cite{Xu2011}.

If the characteristic thickness of the conducting channel $\lambda$ is larger than $t_{sc}$,  the relation (\ref{eqn:Nbulkmax}) is no longer valid to extract the maximum bulk trap density. From the measured subthreshold swing and the device geometry, one can directly asses if $\lambda > t_{sc}$ by rewriting the Debye length $\lambda$ in terms of $S$:
\begin{equation}
\frac{S}{S_{ideal}} < \frac{\varepsilon_{sc}\,t_{ins}}{ t_{sc}\,\varepsilon_{ins}} + 1\,,
\label{eqn:validitycondition}
\end{equation}
where $t_{ins}$ is the thickness and $\varepsilon_{ins}$ the permittivity of the gate dielectric.
Noteworthy, the borderline $\lambda = t_{sc}$ is equivalent to the situation where the trap related effective capacitance $C_{sc}$ equals the geometric capacitance $C_{sc}^{geom}$.

If the Debye length $\lambda$ exceeds the crystal thickness $t_{sc}$, the analysis of $S$ in terms of trap density is slightly modified:
\begin{equation}
	\label{eq:Csc_largelambda}
	C_{sc} = q^2 \left(D_{it} + t_{sc} D_{bulk}\right)
\end{equation}
and accordingly the maximum bulk trap density is
\begin{equation}
	\label{eqn:Nintmax_largelambda}
D_{bulk}^{max} = \frac{C_i}{q^2\, t_{sc}} \left(\frac{q S}{k_B T \ln(10)} - 1 \right)\,.
\end{equation}

After this discussion it is clear that from a physical point of view the as measured subthreshold swing $S$ is not suitable for a direct comparison between different FETs. Sometimes the subthreshold swing multiplied by the gate capacitance has been used for this purpose, but this value still depends on the geometry of the gate dielectric and gives no information about the intrinsic properties of the material. The key quantity to evaluate and compare the quality of the semiconductor and its interface to the dielectric is the capacitance of the transport channel $C_{sc}$ obtained from $S$ according to Eqs. \eqref{eq:S} and \eqref{eq:CscCi}.

\section{Estimation of the trapping energy}

Not only can we extract the density of trap states dominating the subthreshold region, but also their energy relative to the transport level using the following considerations.

The subthreshold current is dominated by diffusion and given by:\cite{Fichtner1979}
\begin{equation}
I_d = q\, W \lambda \, D \,\, \frac{p_{m,s} - p_{m,d}}{L},
\label{eqn:subthr_current}
\end{equation}
where $W$ and $L$ are the width and length of the channel, $\lambda$ is the channel thickness and $D$ is the diffusion coefficient which is connected to the mobility $\mu$ by the Einstein relation $D  = \mu \, k_BT / q$. Since in the subthreshold regime the concentrations of mobile charges near the source, $p_{m,s}$, and the drain, $p_{m,d}$, differ by several orders of magnitude, we set $p_{m,d} = 0$. Furthermore, if the Fermi level lies several $k_B T$ away from the transport level, $p_{m,s}$ is given by Maxwell-Boltzmann statistics:
\begin{equation}
p_{m,s} = N_{band} \, e^{-\frac{\Delta E}{k_B T}}, 
\label{eqn:boltzmann}
\end{equation}
where $N_{band}$ is the number of states in the transport level and $\Delta E = E_F - E_c$ is the Fermi energy relative to the transport level $E_c$.

Again, we consider the two situations where the characteristic channel thickness given by the Debye length $\lambda = \sqrt{\varepsilon_{sc}/ q^2 D_{bulk}}$ is either smaller or larger than the geometric thickness $t_{sc}$ of the semiconductor. For  $\lambda < t_{sc}$, combining Eqs. (\ref{eqn:subthr_current}) and (\ref{eqn:boltzmann}) results in
\begin{equation}
\Delta E = -k_B T \ln \left( \frac{q \, I_d \, L \,}{N_{band} \, W \mu \, k_B T} \, \sqrt{\frac{D_{bulk}^{max}}{\varepsilon_{sc}}} \right)
\label{eqn:trap_depth}
\end{equation}
and the density of bulk traps per unit energy $D_{bulk}^{max}$ can be obtained from the subthreshold swing by using Eq. (\ref{eqn:Nbulkmax}). $I_d$ is the drain current for which the minimal subthreshold swing is reached.
If on the other hand the Debye length exceeds the thickness of the semiconducting layer $(\lambda > t_{sc})$ we obtain
\begin{equation}
\Delta E = -k_B T \ln \left( \frac{I_d \, L \,}{N_{band} \, W \,t_{sc} \,\mu \, k_B T} \right).
\label{eqn:trap_depth2}
\end{equation}

With equation (\ref{eqn:trap_depth}) or (\ref{eqn:trap_depth2}) the Fermi energy and thus the energy of the deep trap states filled upon turn-on can be estimated directly from macroscopic values.

We note in passing that the deepest $\Delta E$ that can be probed is directly given by $k_B T$ and shows a logarithmic dependence on the device geometry and the lowest measurable current (limited by noise level or off scurrent). Therefore, optimizing these parameters gives access to deeper trap states.

\section{Numerical simulation of the subthreshold current}

The analytical method discussed in the previous sections is convenient to estimate the density of deep trap states. Its derivation\cite{RollandDiss}, however, is fairly involved, and it is desirable to assess the validity of this model in detail. We perform a series of numerical simulations of the FET's subthreshold current assuming a wide range of bulk trap densities, and we compare the resulting subthreshold swing with the values predicted by the analytical model (Eq. \ref{eqn:Nbulkmax}).

For the numerical calculations, a new in-house implementation of a FET model is used: It solves the drift- and diffusion equations in the presence of traps, similar to previous simulators.\cite{Oberhoff2007, Scheinert2007, Kim2013a} The differential equations are solved in two dimensions for the entire operating regime of the transistor and the trap DOS can be chosen arbitrarily. Methods to numerically solve these equations are discussed in more detail e.g. in the book \emph{Analysis and Simulation of Semiconductor Devices}.\cite{Selberherr1984}

The drift and diffusion equations for hole-only transport consist of Poisson's equation, the continuity equation, and the definition of the drift and diffusion current density $\vec{J_{p}}$:
\begin{equation}
	\label{eq:poisson}
	\vec{\nabla}^2 \Psi = - \frac{q}{\varepsilon_{sc}} (p_{m} + p_{t})  \hspace{2mm}\,
\end{equation}
\begin{equation}
	\label{eq:continuity}
	\vec{\nabla} \cdot \vec{J_{p}} + q R = - q \frac{\partial{p_{m}}}{\partial{t}}\,
\end{equation}
\begin{equation}
	\label{eq:currents}
	\vec{J_{p}} = -q \left( \mu \, p_{m} \hspace{1.2mm} \vec{\nabla} \Psi + D \, \vec{\nabla} p_{m} \right)\,,
\end{equation}
where $\Psi$ is the electric potential, $p_{m}$ and $p_{t}$ are the mobile and trapped charge carrier densities, respectively, $R$ is the recombination rate (here $R = 0$) and $\mu$ denotes the drift mobility for holes. The diffusion constant $D$ is directly connected to the mobility by the Einstein relation.

This system of non-linear differential equations is discretized into finite differences and solved for the steady-state (i.e. $\partial{p_{m}}/\partial{t} = 0$) in two dimensions, using the Gauss-Newton algorithm.

Dirichlet boundary conditions are used to fix the potential and the charge carrier density at the injecting contacts. The interface to the insulating gate dielectric is determined by the Neumann boundary condition $\vec{J_{p}}\cdot\vec{n}=0$, and the electric fields in the semiconductor and the insulator are connected by Gauss' law,
\begin{equation}
	\label{eq:gauss}
	\varepsilon_{sc} \, \frac{\partial \Psi}{\partial \vec{n}}\bigg|_{sc} - \varepsilon_{ins} \, \frac{\partial \Psi}{\partial \vec{n}}\bigg|_{ins} = Q_{it}\,,
\end{equation}
where $\vec{n}$ is the unit vector orthogonal to the interface plane and $Q_{it}$ is the sheet density of additional charge at the interface. For the simulations discussed here we set $Q_{it}=0$.

The concentration of mobile and trapped charge carriers is determined by a convolution of the density of states with the Fermi-Dirac distribution for holes:
\begin{equation}
	\label{eq:fermi_conv1}
	p_m(E_F,T) = \int {D_{\rm band}(E)\cdot(1 - f(E,E_F,T))} \hspace{1mm} dE
\end{equation}

\begin{equation}
	\label{eq:fermi_conv2}
	p_t(E_F,T) = \int{D_{\rm trap}(E)\cdot (1 - f(E,E_F,T))} \hspace{1mm} dE \,,
\end{equation}
where $D_{band}$ and $D_{trap}$ are the spectral distribution of band-like (mobile) states and traps, respectively. From Eqs. (\ref{eq:fermi_conv1}) and (\ref{eq:fermi_conv2}) a relation $p_m(p_t,T)$ can be calculated for any arbitrary distribution of conducting states and traps.

\begin{figure}
\includegraphics[width=1\columnwidth]{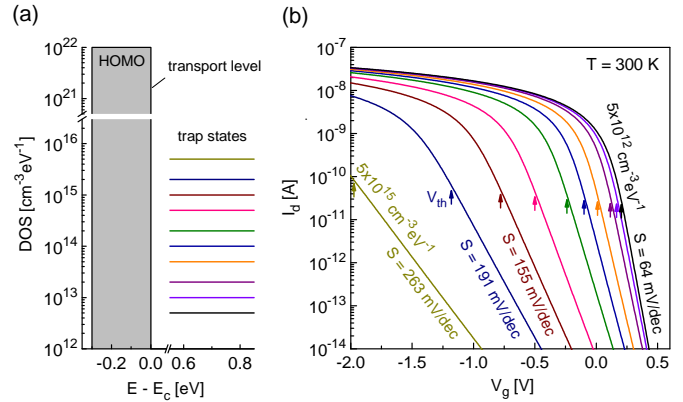}
\caption{ \label{fig:DOS_input_and_transfer} Simulated transfer curves emphasizing the broadening of the subthreshold swing upon increase of the bulk trap DOS. (a) DOS input for the numerical calculations. (b) Simulated transfer curves at $300\,{\rm K}$ for the different bulk trap densities. The transition from the saturation to subthreshold regime is indicated by the threshold voltage (arrows). The subthreshold swing of the rightmost curve corresponds to the value of the measured FET (Fig. \ref{fig:FET_transfer})}
\end{figure}

\begin{figure}
\includegraphics[width=1\columnwidth]{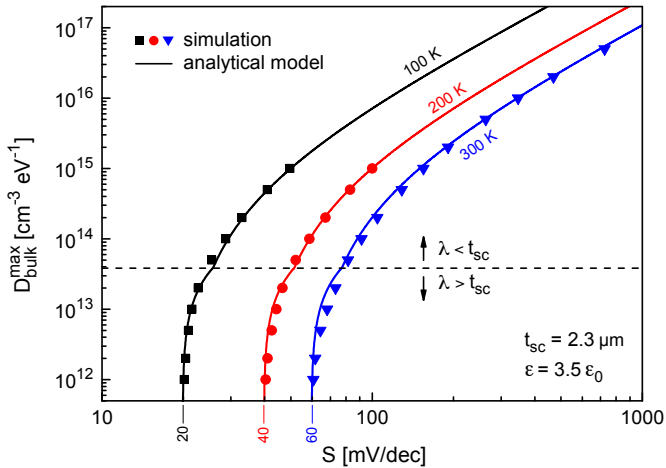}
\caption{ \label{fig:SIM_extraction} Extraction of the bulk trap DOS from the subthreshold swing: Comparison of the simulated subthreshold swing (symbols) and the corresponding prediction by the analytical model (lines) as function of the bulk trap density and for a range of temperatures. In the region below the dashed line, the Debye length $\lambda$ is shorter than the semiconductor thickness, i.e. the characteristic channel thickness is limited by device geometry. For decreasing trap density, the values of $S$ approach their theoretical minimum for the trap-free case, indicated by the (rounded) numbers in the bottom left corner.}
\end{figure}

For this study, the DOS model illustrated in Fig. \ref{fig:DOS_input_and_transfer}(a) was used:  A $0.3 \, {\rm eV}$ wide constant band containing $3 \times 10^{21}$ states per $\rm {cm^{-3}}$ represents the mobile states in the highest occupied molecular orbit level (HOMO). Since we are focussing on the subthreshold region only, we assume for simplicity a constant trap DOS in the energy range relevant at the turn-on voltage. This approximation is reasonable far away from the transport level, as has been measured in previous studies on single-crystal and thin film OFETs, where the trap DOS changes by a factor of $\sim 2$ within a few $k_B T$.\cite{Kalb2010,Kalb2010b,Willa2013} 
We vary the trap density over several orders of magnitude and calculate the transfer characteristics in the subthreshold regime at different temperatures (example in Fig. \ref{fig:DOS_input_and_transfer}(b)). From these curves we take the minimum subthreshold swing $S$ and compare the values in Fig. \ref{fig:SIM_extraction} to the predictions by the analytical model (Eq. \ref{eqn:Nbulkmax} and \ref{eqn:Nintmax_largelambda}, respectively) for the two cases, $\lambda \lessgtr t_{sc}$. First of all, we note the excellent agreement in the entire parameter space. Furthermore, the simulations confirm the need to distinguish between the two ranges (dashed line in Fig. \ref{fig:SIM_extraction}).

\section{Results and Discussion}

The measured low subthreshold swing of $65\,{\rm mV / decade}$ (Figs. \ref{fig:FET_transfer} and \ref{fig:FET_subswing}) is a direct consequence of an exceptionally low density of deep traps (Table \ref{tab:parameters}). Assuming all the measured trap states to be located at the interface, we calculate a $D_{it}^{max}$ of $3 \times 10^{9} {\,\rm cm^{-2}eV^{-1}}$. We compare this density to the number of rubrene molecules at the interface: The lattice parameters of an orthorhombic rubrene crystal are $a = 26.9\,\AA$, $b = 7.2\,\AA$, $c = 14.4\,\AA$,\cite{Jurchescu2006} leading to a density of $9.6 \times 10^{13}$ unit cells per ${\,\rm cm^{2}}$ in the b-c-plane (according to the crystallographic axis-definition) and thus to a molecule density of $3.9 \times 10^{14}\,{\rm cm^{-2}}$. Therefore, the deep trap density per ${\rm eV}$ is less than 1 in $10^5$ molecules. Alternatively, if we we assume all the traps to be distributed throughout the bulk, the same reasoning leads to a density of defects electronically active far away from the transport level of 1 in $10^8$ molecules per $\rm eV$.

Considering the rather limited chemical purity of the source material (98\%), the density of measured defects is remarkably low. This manifests the efficient purification during growth of the crystal by physical vapor transport. Furthermore, the equivalent interface trap density of only $D_{it} = 3 \times 10^{9}~{\rm cm^{-2}eV^{-1}}$ compares very favorably with the best crystalline SiO$_2$/Si interfaces, where $D_{it}$ is in the range of 10$^{10}\,\rm cm^{-2}eV^{-1}$ (to the best of our knowledge\cite{Zhuo2012}).

At turn-on, where the subthreshold swing and thus the trap density has been determined, the Fermi level is calculated to be $\Delta E=0.62\,\rm eV$ above the transport level (Eq. \ref{eqn:trap_depth2}).

 These results indicate that the bulk of the rubrene single crystal is almost defect free, owing to the growth process by physical vapor transport. Also, in contrast to inorganic crystalline semiconductors, no dangling bonds are present on the surface of the van der Waals bonded molecular crystal, which in turn leads to no additional charge traps. Furthermore, the small density of traps per molecule at the surface suggest that any possible intrinsic surface states associated with the termination of a perfectly defect free crystal lattice do not result in a localization of charge carriers. In general, chemical or physical adsorption of molecules to either the Cytop or rubrene is suppressed by their inert surface properties.
 
\setlength{\tabcolsep}{2mm}
\renewcommand{\arraystretch}{1.2}
\begin{table}[tb]
	\begin{center}
	\begin{tabular}{ccc}
		param. & value & unit \\
	\hline
		$T$ & $295$ & K\\
		$S$ & $65\pm2$ &  mV/dec \\
		$\mu_{sat}$ & $13.9$ & ${\rm cm^2/Vs}$ \\
		$n*$ & $1.11\pm0.03$ &\\
		$C_i$ & $4.71\pm 0.09$ & ${\rm nF/cm^2}$ \\
		$C_{sc}$ & $ 0.52 \pm 0.16$ & ${\rm nF/cm^2}$ \\
		$\varepsilon_{\rm sc,rel}$ & $3.5$ & \\
		$D_{bulk}^{max}$ & $(1.3\pm0.4) \times 10^{13}$ & ${\rm cm^{-3}eV^{-1}}$\\
		$D_{it}^{max}$ & $(3\pm1) \times 10^{9}$ & ${\rm cm^{-2}eV^{-1}}$\\
		$\Delta E$ & $ 0.62 $ & eV\\
	\end{tabular}
	\caption{Results and parameters relevant for the extraction deep trap density in the measured rubrene single crystal FET.}
	\label{tab:parameters}
	\end{center}
\end{table}

From the methodic aspect, these measurements are of particular interest because the calculated Debye length of $\sim 4\,\rm {\mu m}$ places this device in the parameter space where the channel thickness at turn-on is limited by the crystal thickness ($2.3\, {\rm \mu m}$). The comparison of the analytical DOS extraction method with the numerical simulations reveals good general agreement in a broad range of bulk trap densities and temperatures (Fig. \ref{fig:SIM_extraction}). Not surprisingly, the small deviations are most pronounced near the borderline at which we distinguish between the situations of a characteristic channel length shorter or larger than the actual semiconductor thickness. However, even in this region the difference in $D_{bulk}$ is at most half an order of magnitude, indicating that this method is well suited also for OFETs with a nearly trap-free interface and bulk, or with a very thin semiconducting layer.

\section{Conclusion}

The rubrene FETs in this study have an extremely low density of deep trap states, indicating the high quality of the semiconductor. Here, this quality is seen in an unprecedentedly low subthreshold swing of $S = 65\,{\rm mV/decade}$ at room temperature, which lies remarkably close to the theoretical trap free limit at $58.5\,{\rm mV/decade}$.

An analytical method is shown to be well suited to experimentally access the density of deep bulk and interface traps from the subthreshold region of FET transfer curves. This method was verified by a comparison to numerical simulations of the subthreshold current. A novel way to estimate the trapping energies was presented which additionally provides the energy range dominating the subthreshold region.

With this in-depth analysis we estimate a trap density of $D_{bulk} = 1 \times 10^{13}~{\rm cm^{-3}eV^{-1}}$, or equivalently $D_{it} = 3 \times 10^{9}~{\rm cm^{-2}eV^{-1}}$, at $0.62~~{\rm eV}$ above the transport level. Thus, the deep trap densities in the best organic single crystal FETs can be lower than in the most advanced crystalline SiO$_2$/Si transistor ($D_{it}^{Si} = 10^{10}\,\rm cm^{-2}eV^{-1}$.\cite{Zhuo2012}). This may come as a surprise considering the flip-crystal fabrication of the OFETs which does not involve any UHV equipment.

In fact, the low trap density is an immediate consequence of the electronically inert and chemically stable surface of the van der Waals bonded molecular organic semiconductors as well as their intrinsically trap-free interface with the gate dielectric. These highly pure crystals are a promising base for further studies of the intrinsic electronic properties of organic semiconductors.
	
%


\begin{thebibliography}{38}%
\makeatletter
\providecommand \@ifxundefined [1]{%
 \@ifx{#1\undefined}
}%
\providecommand \@ifnum [1]{%
 \ifnum #1\expandafter \@firstoftwo
 \else \expandafter \@secondoftwo
 \fi
}%
\providecommand \@ifx [1]{%
 \ifx #1\expandafter \@firstoftwo
 \else \expandafter \@secondoftwo
 \fi
}%
\providecommand \natexlab [1]{#1}%
\providecommand \enquote  [1]{``#1''}%
\providecommand \bibnamefont  [1]{#1}%
\providecommand \bibfnamefont [1]{#1}%
\providecommand \citenamefont [1]{#1}%
\providecommand \href@noop [0]{\@secondoftwo}%
\providecommand \href [0]{\begingroup \@sanitize@url \@href}%
\providecommand \@href[1]{\@@startlink{#1}\@@href}%
\providecommand \@@href[1]{\endgroup#1\@@endlink}%
\providecommand \@sanitize@url [0]{\catcode `\\12\catcode `\$12\catcode
  `\&12\catcode `\#12\catcode `\^12\catcode `\_12\catcode `\%12\relax}%
\providecommand \@@startlink[1]{}%
\providecommand \@@endlink[0]{}%
\providecommand \url  [0]{\begingroup\@sanitize@url \@url }%
\providecommand \@url [1]{\endgroup\@href {#1}{\urlprefix }}%
\providecommand \urlprefix  [0]{URL }%
\providecommand \Eprint [0]{\href }%
\providecommand \doibase [0]{http://dx.doi.org/}%
\providecommand \selectlanguage [0]{\@gobble}%
\providecommand \bibinfo  [0]{\@secondoftwo}%
\providecommand \bibfield  [0]{\@secondoftwo}%
\providecommand \translation [1]{[#1]}%
\providecommand \BibitemOpen [0]{}%
\providecommand \bibitemStop [0]{}%
\providecommand \bibitemNoStop [0]{.\EOS\space}%
\providecommand \EOS [0]{\spacefactor3000\relax}%
\providecommand \BibitemShut  [1]{\csname bibitem#1\endcsname}%
\let\auto@bib@innerbib\@empty

\bibitem [{\citenamefont {Bussolotti2013}, \citenamefont {Kurokawa},\ and\
  \citenamefont {Bussolotti}(1983)}]{Bussolotti2013}%
  \BibitemOpen
  \bibfield  {author} {\bibinfo {author} {\bibfnamefont {F.}~\bibnamefont
  {Bussolotti}}, \bibinfo {author} {\bibfnamefont {S.}~\bibnamefont {Kera}}, \bibinfo {author} {\bibfnamefont {K.}~\bibnamefont {Kudo}}, \bibinfo {author} {\bibfnamefont {A.}~\bibnamefont {Kahn}} \
  and\ \bibinfo {author} {\bibfnamefont {N.}~\bibnamefont {Ueno}},\ } {\bibfield  {journal} {\bibinfo  {journal}
  {Physical Review Letters}\ }\textbf {\bibinfo {volume} {110}},\ \bibinfo
  {pages} {267602} (\bibinfo {year} {2013})}\BibitemShut {NoStop}%
\bibitem [{\citenamefont {Mishchenko2012}, \citenamefont {Mishchenko}(2012)}]{Mishchenko2012}%
  \BibitemOpen
  \bibfield  {author} {\bibinfo {author} {\bibfnamefont {A.S.}~\bibnamefont
  {Mishchenko}}, \bibinfo {author} {\bibfnamefont {H.}~\bibnamefont {Matsui}}, 
  and\ \bibinfo {author} {\bibfnamefont {T.}~\bibnamefont {Hasegawa}},\ } {\bibfield  {journal} {\bibinfo  {journal}
  {Physical Review B}\ }\textbf {\bibinfo {volume} {85}},\ \bibinfo
  {pages} {085211} (\bibinfo {year} {2012})}\BibitemShut {NoStop}%
\bibitem [{\citenamefont {Powell1992}, \citenamefont {Powell}(1992)}]{Powell1992}%
  \BibitemOpen
  \bibfield  {author} {\bibinfo {author} {\bibfnamefont {M.J.}~\bibnamefont
  {Powell}}, \bibinfo {author} {\bibfnamefont {C}~\bibnamefont {vanBerkel}}, 
  and\ \bibinfo {author} {\bibfnamefont {A.R.}~\bibnamefont {Franklin}}, \bibinfo {author} {\bibfnamefont {S.C.}~\bibnamefont {Deane}}, \bibinfo {author} {\bibfnamefont {W.I.}~\bibnamefont {Milne}},\ } {\bibfield  {journal} {\bibinfo  {journal}
  {Physical Review B}\ }\textbf {\bibinfo {volume} {45}},\ \bibinfo
  {pages} {4160} (\bibinfo {year} {1992})}\BibitemShut {NoStop}%
\bibitem [{\citenamefont {Kalb}\ and\ \citenamefont
  {Batlogg}(2010)}]{Kalb2010}%
  \BibitemOpen
  \bibfield  {author} {\bibinfo {author} {\bibfnamefont {W.~L.}\ \bibnamefont
  {Kalb}}\ and\ \bibinfo {author} {\bibfnamefont {B.}~\bibnamefont {Batlogg}},\
  }\href {\doibase 10.1103/PhysRevB.81.035327} {\bibfield  {journal} {\bibinfo
  {journal} {Phys. Rev. B}\ }\textbf {\bibinfo {volume} {81}},\ \bibinfo
  {pages} {035327} (\bibinfo {year} {2010})}\BibitemShut {NoStop}%
\bibitem [{\citenamefont {Ebisawa}, \citenamefont {Kurokawa},\ and\
  \citenamefont {Nara}(1983)}]{Ebisawa1983}%
  \BibitemOpen
  \bibfield  {author} {\bibinfo {author} {\bibfnamefont {F.}~\bibnamefont
  {Ebisawa}}, \bibinfo {author} {\bibfnamefont {T.}~\bibnamefont {Kurokawa}}, \
  and\ \bibinfo {author} {\bibfnamefont {S.}~\bibnamefont {Nara}},\ }\href
  {\doibase 10.1063/1.332488} {\bibfield  {journal} {\bibinfo  {journal}
  {Journal of Applied Physics}\ }\textbf {\bibinfo {volume} {54}},\ \bibinfo
  {pages} {3255} (\bibinfo {year} {1983})}\BibitemShut {NoStop}%
\bibitem [{\citenamefont {Horowitz}, \citenamefont {Garnier},\ and\
  \citenamefont {Yassar}(1996)}]{Horowitz1996}%
  \BibitemOpen
  \bibfield  {author} {\bibinfo {author} {\bibfnamefont {G.}~\bibnamefont
  {Horowitz}}, \bibinfo {author} {\bibfnamefont {F.}~\bibnamefont {Garnier}}, \
  and\ \bibinfo {author} {\bibfnamefont {A.}~\bibnamefont {Yassar}},\ }\href
  {http://onlinelibrary.wiley.com/doi/10.1002/adma.19960080109/full} {\bibfield
   {journal} {\bibinfo  {journal} {Advanced Materials}\ }\textbf {\bibinfo
  {volume} {8}},\ \bibinfo {pages} {52} (\bibinfo {year} {1996})}\BibitemShut
  {NoStop}%
\bibitem [{\citenamefont {Takeya}\ \emph {et~al.}(2007)\citenamefont {Takeya},
  \citenamefont {Yamagishi}, \citenamefont {Tominari}, \citenamefont
  {Hirahara}, \citenamefont {Nakazawa}, \citenamefont {Nishikawa},
  \citenamefont {Kawase}, \citenamefont {Shimoda},\ and\ \citenamefont
  {Ogawa}}]{Takeya2007}%
  \BibitemOpen
  \bibfield  {author} {\bibinfo {author} {\bibfnamefont {J.}~\bibnamefont
  {Takeya}}, \bibinfo {author} {\bibfnamefont {M.}~\bibnamefont {Yamagishi}},
  \bibinfo {author} {\bibfnamefont {Y.}~\bibnamefont {Tominari}}, \bibinfo
  {author} {\bibfnamefont {R.}~\bibnamefont {Hirahara}}, \bibinfo {author}
  {\bibfnamefont {Y.}~\bibnamefont {Nakazawa}}, \bibinfo {author}
  {\bibfnamefont {T.}~\bibnamefont {Nishikawa}}, \bibinfo {author}
  {\bibfnamefont {T.}~\bibnamefont {Kawase}}, \bibinfo {author} {\bibfnamefont
  {T.}~\bibnamefont {Shimoda}}, \ and\ \bibinfo {author} {\bibfnamefont
  {S.}~\bibnamefont {Ogawa}},\ }\href {\doibase 10.1063/1.2711393} {\bibfield
  {journal} {\bibinfo  {journal} {Applied Physics Letters}\ }\textbf {\bibinfo
  {volume} {90}},\ \bibinfo {pages} {102120} (\bibinfo {year}
  {2007})}\BibitemShut {NoStop}%
\bibitem [{\citenamefont {Lin}\ \emph {et~al.}(1997)\citenamefont {Lin},
  \citenamefont {Member}, \citenamefont {Gundlach}, \citenamefont {Nelson},\
  and\ \citenamefont {Jackson}}]{Lin1997}%
  \BibitemOpen
  \bibfield  {author} {\bibinfo {author} {\bibfnamefont {Y.-y.}\ \bibnamefont
  {Lin}}, \bibinfo {author} {\bibfnamefont {S.}~\bibnamefont {Member}},
  \bibinfo {author} {\bibfnamefont {D.~J.}\ \bibnamefont {Gundlach}}, \bibinfo
  {author} {\bibfnamefont {S.~F.}\ \bibnamefont {Nelson}}, \ and\ \bibinfo
  {author} {\bibfnamefont {T.~N.}\ \bibnamefont {Jackson}},\ }\href@noop {}
  {\bibfield  {journal} {\bibinfo  {journal} {IEEE Transactions On Electron
  Devices}\ }\textbf {\bibinfo {volume} {44}},\ \bibinfo {pages} {1325}
  (\bibinfo {year} {1997})}\BibitemShut {NoStop}%
\bibitem [{\citenamefont {Sirringhaus}\ \emph {et~al.}(1999)\citenamefont
  {Sirringhaus}, \citenamefont {Tessler}, \citenamefont {Thomas}, \citenamefont
  {Brown},\ and\ \citenamefont {Friend}}]{Sirringhaus1999}%
  \BibitemOpen
  \bibfield  {author} {\bibinfo {author} {\bibfnamefont {H.}~\bibnamefont
  {Sirringhaus}}, \bibinfo {author} {\bibfnamefont {N.}~\bibnamefont
  {Tessler}}, \bibinfo {author} {\bibfnamefont {D.~S.}\ \bibnamefont {Thomas}},
  \bibinfo {author} {\bibfnamefont {P.~J.}\ \bibnamefont {Brown}}, \ and\
  \bibinfo {author} {\bibfnamefont {R.~H.}\ \bibnamefont {Friend}},\
  }\href@noop {} {\bibfield  {journal} {\bibinfo  {journal} {Adv. Solid State
  Phys.}\ }\textbf {\bibinfo {volume} {39}},\ \bibinfo {pages} {101} (\bibinfo
  {year} {1999})}\BibitemShut {NoStop}%
\bibitem [{\citenamefont {Laudise}\ \emph {et~al.}(1998)\citenamefont
  {Laudise}, \citenamefont {Kloc}, \citenamefont {Simpkins},\ and\
  \citenamefont {Siegrist}}]{Laudise1998}%
  \BibitemOpen
  \bibfield  {author} {\bibinfo {author} {\bibfnamefont {R. A.}~\bibnamefont
  {Laudise}}, \bibinfo {author} {\bibfnamefont {C.}~\bibnamefont {Kloc}},
  \bibinfo {author} {\bibfnamefont {P.}~\bibnamefont {Simpkins}}, \ and\
  \bibinfo {author} {\bibfnamefont {T.}~\bibnamefont {Siegrist}},\ }\href
  {\doibase 10.1016/S0022-0248(98)00034-7} {\bibfield  {journal} {\bibinfo
  {journal} {Journal of Crystal Growth}\ }\textbf {\bibinfo {volume} {187}},\
  \bibinfo {pages} {449} (\bibinfo {year} {1998})}\BibitemShut {NoStop}%
\bibitem [{\citenamefont {Podzorov}\ \emph {et~al.}(2003)\citenamefont
  {Podzorov}, \citenamefont {Sysoev}, \citenamefont {Loginova}, \citenamefont
  {Pudalov},\ and\ \citenamefont {Gershenson}}]{Podzorov2003a}%
  \BibitemOpen
  \bibfield  {author} {\bibinfo {author} {\bibfnamefont {V.}~\bibnamefont
  {Podzorov}}, \bibinfo {author} {\bibfnamefont {S.~E.}\ \bibnamefont
  {Sysoev}}, \bibinfo {author} {\bibfnamefont {E.}~\bibnamefont {Loginova}},
  \bibinfo {author} {\bibfnamefont {V.~M.}\ \bibnamefont {Pudalov}}, \ and\
  \bibinfo {author} {\bibfnamefont {M.~E.}\ \bibnamefont {Gershenson}},\ }\href
  {\doibase 10.1063/1.1622799} {\bibfield  {journal} {\bibinfo  {journal}
  {Applied Physics Letters}\ }\textbf {\bibinfo {volume} {83}},\ \bibinfo
  {pages} {3504} (\bibinfo {year} {2003})}\BibitemShut {NoStop}%
\bibitem [{\citenamefont {Podzorov}\ \emph {et~al.}(2004)\citenamefont
  {Podzorov}, \citenamefont {Menard}, \citenamefont {Borissov}, \citenamefont
  {Kiryukhin}, \citenamefont {Rogers},\ and\ \citenamefont
  {Gershenson}}]{Podzorov2004}%
  \BibitemOpen
  \bibfield  {author} {\bibinfo {author} {\bibfnamefont {V.}~\bibnamefont
  {Podzorov}}, \bibinfo {author} {\bibfnamefont {E.}~\bibnamefont {Menard}},
  \bibinfo {author} {\bibfnamefont {A.}~\bibnamefont {Borissov}}, \bibinfo
  {author} {\bibfnamefont {V.}~\bibnamefont {Kiryukhin}}, \bibinfo {author}
  {\bibfnamefont {J.~A.}\ \bibnamefont {Rogers}}, \ and\ \bibinfo {author}
  {\bibfnamefont {M.~E.}\ \bibnamefont {Gershenson}},\ }\href {\doibase
  10.1103/PhysRevLett.93.086602} {\bibfield  {journal} {\bibinfo  {journal}
  {Physical Review Letters}\ }\textbf {\bibinfo {volume} {93}},\ \bibinfo
  {pages} {086602} (\bibinfo {year} {2004})}\BibitemShut {NoStop}%
\bibitem [{\citenamefont {Xie}\ \emph {et~al.}(2013)\citenamefont {Xie},
  \citenamefont {Willa}, \citenamefont {Wu}, \citenamefont {Hausermann},
  \citenamefont {Takimiya}, \citenamefont {Batlogg},\ and\ \citenamefont
  {Frisbie}}]{Xie2013}%
  \BibitemOpen
  \bibfield  {author} {\bibinfo {author} {\bibfnamefont {W.}~\bibnamefont
  {Xie}}, \bibinfo {author} {\bibfnamefont {K.}~\bibnamefont {Willa}}, \bibinfo
  {author} {\bibfnamefont {Y.}~\bibnamefont {Wu}}, \bibinfo {author}
  {\bibfnamefont {R.}~\bibnamefont {Hausermann}}, \bibinfo {author}
  {\bibfnamefont {K.}~\bibnamefont {Takimiya}}, \bibinfo {author}
  {\bibfnamefont {B.}~\bibnamefont {Batlogg}}, \ and\ \bibinfo {author}
  {\bibfnamefont {C.~D.}\ \bibnamefont {Frisbie}},\ }\href {\doibase
  10.1002/adma.201300886} {\bibfield  {journal} {\bibinfo  {journal} {Advanced
  Materials}\ }\textbf {\bibinfo {volume} {25}},\ \bibinfo {pages} {3478}
  (\bibinfo {year} {2013})}\BibitemShut {NoStop}%
\bibitem [{\citenamefont {Kalb}\ \emph {et~al.}(2007)\citenamefont {Kalb},
  \citenamefont {Mathis}, \citenamefont {Haas}, \citenamefont {Stassen},\ and\
  \citenamefont {Batlogg}}]{Kalb2007}%
  \BibitemOpen
  \bibfield  {author} {\bibinfo {author} {\bibfnamefont {W.~L.}\ \bibnamefont
  {Kalb}}, \bibinfo {author} {\bibfnamefont {T.}~\bibnamefont {Mathis}},
  \bibinfo {author} {\bibfnamefont {S.}~\bibnamefont {Haas}}, \bibinfo {author}
  {\bibfnamefont {A.~F.}\ \bibnamefont {Stassen}}, \ and\ \bibinfo {author}
  {\bibfnamefont {B.}~\bibnamefont {Batlogg}},\ }\href@noop {} {\bibfield
  {journal} {\bibinfo  {journal} {Applied Physics Letters}\ }\textbf {\bibinfo
  {volume} {90}},\ \bibinfo {pages} {92104} (\bibinfo {year}
  {2007})}\BibitemShut {NoStop}%
\bibitem [{\citenamefont {Willa}\ \emph {et~al.}(2013)\citenamefont {Willa},
  \citenamefont {H\"{a}usermann}, \citenamefont {Mathis}, \citenamefont
  {Facchetti}, \citenamefont {Chen},\ and\ \citenamefont
  {Batlogg}}]{Willa2013}%
  \BibitemOpen
  \bibfield  {author} {\bibinfo {author} {\bibfnamefont {K.}~\bibnamefont
  {Willa}}, \bibinfo {author} {\bibfnamefont {R.}~\bibnamefont
  {H\"{a}usermann}}, \bibinfo {author} {\bibfnamefont {T.}~\bibnamefont
  {Mathis}}, \bibinfo {author} {\bibfnamefont {A.}~\bibnamefont {Facchetti}},
  \bibinfo {author} {\bibfnamefont {Z.}~\bibnamefont {Chen}}, \ and\ \bibinfo
  {author} {\bibfnamefont {B.}~\bibnamefont {Batlogg}},\ }\href {\doibase
  10.1063/1.4798610} {\bibfield  {journal} {\bibinfo  {journal} {J. Appl.
  Phys.}\ }\textbf {\bibinfo {volume} {113}},\ \bibinfo {pages} {133707}
  (\bibinfo {year} {2013})}\BibitemShut {NoStop}%
\bibitem [{\citenamefont {Zschieschang}\ \emph {et~al.}(2012)\citenamefont
  {Zschieschang}, \citenamefont {Kang}, \citenamefont {Takimiya}, \citenamefont
  {Sekitani}, \citenamefont {Someya}, \citenamefont {Canzler}, \citenamefont
  {Werner}, \citenamefont {Blochwitz-Nimoth},\ and\ \citenamefont
  {Klauk}}]{Zschieschang2012}%
  \BibitemOpen
  \bibfield  {author} {\bibinfo {author} {\bibfnamefont {U.}~\bibnamefont
  {Zschieschang}}, \bibinfo {author} {\bibfnamefont {M.~J.}\ \bibnamefont
  {Kang}}, \bibinfo {author} {\bibfnamefont {K.}~\bibnamefont {Takimiya}},
  \bibinfo {author} {\bibfnamefont {T.}~\bibnamefont {Sekitani}}, \bibinfo
  {author} {\bibfnamefont {T.}~\bibnamefont {Someya}}, \bibinfo {author}
  {\bibfnamefont {T.~W.}\ \bibnamefont {Canzler}}, \bibinfo {author}
  {\bibfnamefont {A.}~\bibnamefont {Werner}}, \bibinfo {author} {\bibfnamefont
  {J.}~\bibnamefont {Blochwitz-Nimoth}}, \ and\ \bibinfo {author}
  {\bibfnamefont {H.}~\bibnamefont {Klauk}},\ }\href {\doibase
  10.1039/c1jm14917b} {\bibfield  {journal} {\bibinfo  {journal} {Journal of
  Materials Chemistry}\ }\textbf {\bibinfo {volume} {22}},\ \bibinfo {pages}
  {4273} (\bibinfo {year} {2012})}\BibitemShut {NoStop}%
\bibitem [{\citenamefont {Ma}\ \emph {et~al.}(2008)\citenamefont {Ma},
  \citenamefont {Acton}, \citenamefont {Ting}, \citenamefont {Ka},
  \citenamefont {Yip}, \citenamefont {Tucker}, \citenamefont {Schofield},\ and\
  \citenamefont {Jen}}]{Ma2008}%
  \BibitemOpen
  \bibfield  {author} {\bibinfo {author} {\bibfnamefont {H.}~\bibnamefont
  {Ma}}, \bibinfo {author} {\bibfnamefont {O.}~\bibnamefont {Acton}}, \bibinfo
  {author} {\bibfnamefont {G.}~\bibnamefont {Ting}}, \bibinfo {author}
  {\bibfnamefont {J.~W.}\ \bibnamefont {Ka}}, \bibinfo {author} {\bibfnamefont
  {H.-L.}\ \bibnamefont {Yip}}, \bibinfo {author} {\bibfnamefont
  {N.}~\bibnamefont {Tucker}}, \bibinfo {author} {\bibfnamefont
  {R.}~\bibnamefont {Schofield}}, \ and\ \bibinfo {author} {\bibfnamefont
  {A.~K.-Y.}\ \bibnamefont {Jen}},\ }\href {\doibase 10.1063/1.2857502}
  {\bibfield  {journal} {\bibinfo  {journal} {Applied Physics Letters}\
  }\textbf {\bibinfo {volume} {92}},\ \bibinfo {pages} {113303} (\bibinfo
  {year} {2008})}\BibitemShut {NoStop}%
\bibitem [{\citenamefont {Liu}\ \emph {et~al.}(2009)\citenamefont {Liu},
  \citenamefont {Oh}, \citenamefont {Roberts}, \citenamefont {Wei},
  \citenamefont {Paul}, \citenamefont {Okajima}, \citenamefont {Nishi},\ and\
  \citenamefont {Bao}}]{Liu2009}%
  \BibitemOpen
  \bibfield  {author} {\bibinfo {author} {\bibfnamefont {Z.}~\bibnamefont
  {Liu}}, \bibinfo {author} {\bibfnamefont {J.~H.}\ \bibnamefont {Oh}},
  \bibinfo {author} {\bibfnamefont {M.~E.}\ \bibnamefont {Roberts}}, \bibinfo
  {author} {\bibfnamefont {P.}~\bibnamefont {Wei}}, \bibinfo {author}
  {\bibfnamefont {B.~C.}\ \bibnamefont {Paul}}, \bibinfo {author}
  {\bibfnamefont {M.}~\bibnamefont {Okajima}}, \bibinfo {author} {\bibfnamefont
  {Y.}~\bibnamefont {Nishi}}, \ and\ \bibinfo {author} {\bibfnamefont
  {Z.}~\bibnamefont {Bao}},\ }\href {\doibase 10.1063/1.3133902} {\bibfield
  {journal} {\bibinfo  {journal} {Applied Physics Letters}\ }\textbf {\bibinfo
  {volume} {94}},\ \bibinfo {pages} {203301} (\bibinfo {year}
  {2009})}\BibitemShut {NoStop}%
\bibitem [{\citenamefont {Walser}\ \emph {et~al.}(2009)\citenamefont {Walser},
  \citenamefont {Kalb}, \citenamefont {Mathis},\ and\ \citenamefont
  {Batlogg}}]{Walser2009a}%
  \BibitemOpen
  \bibfield  {author} {\bibinfo {author} {\bibfnamefont {M.~P.}\ \bibnamefont
  {Walser}}, \bibinfo {author} {\bibfnamefont {W.~L.}\ \bibnamefont {Kalb}},
  \bibinfo {author} {\bibfnamefont {T.}~\bibnamefont {Mathis}}, \ and\ \bibinfo
  {author} {\bibfnamefont {B.}~\bibnamefont {Batlogg}},\ }\href
  {http://dx.doi.org/10.1063/1.3267055} {\bibfield  {journal} {\bibinfo
  {journal} {Applied Physics Letters}\ }\textbf {\bibinfo {volume} {95}},\
  \bibinfo {pages} {233301} (\bibinfo {year} {2009})}\BibitemShut {NoStop}%
\bibitem [{\citenamefont {Park}\ \emph {et~al.}(2012)\citenamefont {Park},
  \citenamefont {Lee}, \citenamefont {Lee}, \citenamefont {Lee}, \citenamefont
  {Lee}, \citenamefont {Yoon}, \citenamefont {Sung},\ and\ \citenamefont
  {Im}}]{Park2012}%
  \BibitemOpen
  \bibfield  {author} {\bibinfo {author} {\bibfnamefont {J.~H.}\ \bibnamefont
  {Park}}, \bibinfo {author} {\bibfnamefont {H.~S.}\ \bibnamefont {Lee}},
  \bibinfo {author} {\bibfnamefont {J.}~\bibnamefont {Lee}}, \bibinfo {author}
  {\bibfnamefont {K.}~\bibnamefont {Lee}}, \bibinfo {author} {\bibfnamefont
  {G.}~\bibnamefont {Lee}}, \bibinfo {author} {\bibfnamefont {K.~H.}\
  \bibnamefont {Yoon}}, \bibinfo {author} {\bibfnamefont {M.~M.}\ \bibnamefont
  {Sung}}, \ and\ \bibinfo {author} {\bibfnamefont {S.}~\bibnamefont {Im}},\
  }\href {\doibase 10.1039/c2cp41544e} {\bibfield  {journal} {\bibinfo
  {journal} {Physical chemistry chemical physics : PCCP}\ }\textbf {\bibinfo
  {volume} {14}},\ \bibinfo {pages} {14202} (\bibinfo {year}
  {2012})}\BibitemShut {NoStop}%
\bibitem [{\citenamefont {Ono}\ \emph {et~al.}(2013)\citenamefont {Ono},
  \citenamefont {Hausermann}, \citenamefont {Chiba}, \citenamefont {Shimamura},
  \citenamefont {Ono},\ and\ \citenamefont {Batlogg}}]{Ono2013}%
  \BibitemOpen
  \bibfield  {author} {\bibinfo {author} {\bibfnamefont {S.}~\bibnamefont
  {Ono}}, \bibinfo {author} {\bibfnamefont {R.}~\bibnamefont {Hausermann}},
  \bibinfo {author} {\bibfnamefont {D.}~\bibnamefont {Chiba}}, \bibinfo
  {author} {\bibfnamefont {K.}~\bibnamefont {Shimamura}}, \bibinfo {author}
  {\bibfnamefont {T.}~\bibnamefont {Ono}}, \ and\ \bibinfo {author}
  {\bibfnamefont {B.}~\bibnamefont {Batlogg}},\ }\href@noop {} {\bibfield
  {journal} {\bibinfo  {journal} {Applied Physics Letters}\ }\textbf {\bibinfo {volume} {104}}, \ \bibinfo {pages} {013307} (\bibinfo {year}
  {2014})}\BibitemShut {NoStop}%
\bibitem [{\citenamefont {Takeya}\ \emph {et~al.}(2003)\citenamefont {Takeya},
  \citenamefont {Goldmann}, \citenamefont {Haas}, \citenamefont {Pernstich},
  \citenamefont {Ketterer},\ and\ \citenamefont {Batlogg}}]{Takeya2003a}%
  \BibitemOpen
  \bibfield  {author} {\bibinfo {author} {\bibfnamefont {J.}~\bibnamefont
  {Takeya}}, \bibinfo {author} {\bibfnamefont {C.}~\bibnamefont {Goldmann}},
  \bibinfo {author} {\bibfnamefont {S.}~\bibnamefont {Haas}}, \bibinfo {author}
  {\bibfnamefont {K.~P.}\ \bibnamefont {Pernstich}}, \bibinfo {author}
  {\bibfnamefont {B.}~\bibnamefont {Ketterer}}, \ and\ \bibinfo {author}
  {\bibfnamefont {B.}~\bibnamefont {Batlogg}},\ }\href {\doibase
  10.1063/1.1618919} {\bibfield  {journal} {\bibinfo  {journal} {Journal of
  Applied Physics}\ }\textbf {\bibinfo {volume} {94}},\ \bibinfo {pages} {5800}
  (\bibinfo {year} {2003})}\BibitemShut {NoStop}%
\bibitem [{\citenamefont {de Boer}\ \emph {et~al.}(2003)\citenamefont {de Boer},
  \citenamefont {Klapwijk},\ and\ \citenamefont {Morpurgo}}]{deBoer2003}%
  \BibitemOpen
  \bibfield  {author} {\bibinfo {author} {\bibfnamefont {R. W. I.}~\bibnamefont
  {de Boer}}, \bibinfo {author} {\bibfnamefont {T. M.}~\bibnamefont {Klapwijk}} \ and\ \bibinfo {author}
  {\bibfnamefont {A. F.}~\bibnamefont {Morpurgo}},\ }\href {\doibase 10.1063/1.1629144} {\bibfield  {journal}
  {\bibinfo  {journal} {Applied Physics Letters}\ }\textbf {\bibinfo {volume}
  {83}},\ \bibinfo {pages} {4345} (\bibinfo {year} {2003})}\BibitemShut
  {NoStop}%
\bibitem [{\citenamefont {Bobbert}\ \emph {et~al.}(2012)\citenamefont
  {Bobbert}, \citenamefont {Sharma}, \citenamefont {Mathijssen}, \citenamefont
  {Kemerink},\ and\ \citenamefont {de~Leeuw}}]{Bobbert2012}%
  \BibitemOpen
  \bibfield  {author} {\bibinfo {author} {\bibfnamefont {P.~A.}\ \bibnamefont
  {Bobbert}}, \bibinfo {author} {\bibfnamefont {A.}~\bibnamefont {Sharma}},
  \bibinfo {author} {\bibfnamefont {S.~G.~J.}\ \bibnamefont {Mathijssen}},
  \bibinfo {author} {\bibfnamefont {M.}~\bibnamefont {Kemerink}}, \ and\
  \bibinfo {author} {\bibfnamefont {D.~M.}\ \bibnamefont {de~Leeuw}},\ }\href
  {\doibase 10.1002/adma.201104580} {\bibfield  {journal} {\bibinfo  {journal}
  {Advanced materials (Deerfield Beach, Fla.)}\ }\textbf {\bibinfo {volume}
  {24}},\ \bibinfo {pages} {1146} (\bibinfo {year} {2012})}\BibitemShut
  {NoStop}%
\bibitem [{\citenamefont {H\"{a}usermann}\ and\ \citenamefont
  {Batlogg}(2011)}]{Hausermann2011}%
  \BibitemOpen
  \bibfield  {author} {\bibinfo {author} {\bibfnamefont {R.}~\bibnamefont
  {H\"{a}usermann}}\ and\ \bibinfo {author} {\bibfnamefont {B.}~\bibnamefont
  {Batlogg}},\ }\href {\doibase 10.1063/1.3628297} {\bibfield  {journal}
  {\bibinfo  {journal} {Applied Physics Letters}\ }\textbf {\bibinfo {volume}
  {99}},\ \bibinfo {pages} {083303} (\bibinfo {year} {2011})}\BibitemShut
  {NoStop}%
\bibitem [{\citenamefont {Pernstich}(2012)}]{Pernstich2012}%
  \BibitemOpen
  \bibfield  {author} {\bibinfo {author} {\bibfnamefont {K.}~\bibnamefont
  {Pernstich}},\ }\href
  {http://nvlpubs.nist.gov/nistpubs/jres/117/jres.117.010.pdf} {\bibfield
  {journal} {\bibinfo  {journal} {Journal of Research of the National Institute
  of Standards and Technology}\ }\textbf {\bibinfo {volume} {117}},\ \bibinfo
  {pages} {176} (\bibinfo {year} {2012})}\BibitemShut {NoStop}%
\bibitem [{\citenamefont {Sze}\ and\ \citenamefont
  {Ng}(2006)}]{Sze1981physics}%
  \BibitemOpen
  \bibfield  {author} {\bibinfo {author} {\bibfnamefont {S.}~\bibnamefont
  {Sze}}\ and\ \bibinfo {author} {\bibfnamefont {K.}~\bibnamefont {Ng}},\
  }\href {http://books.google.ch/books?id=o4unkmHBHb8C} {\emph {\bibinfo
  {title} {Physics of Semiconductor Devices}}}\ (\bibinfo  {publisher}
  {Wiley},\ \bibinfo {year} {2006})\BibitemShut {NoStop}%
\bibitem [{\citenamefont {Colinge}\ and\ \citenamefont
  {Colinge}(2002)}]{colinge2002physics}%
  \BibitemOpen
  \bibfield  {author} {\bibinfo {author} {\bibfnamefont {J.}~\bibnamefont
  {Colinge}}\ and\ \bibinfo {author} {\bibfnamefont {C.}~\bibnamefont
  {Colinge}},\ }\href {http://books.google.ch/books?id=ZcDE-ENKh2gC} {\emph
  {\bibinfo {title} {Physics of Semiconductor Devices}}},\ Physics of
  semiconductor devices\ (\bibinfo  {publisher} {Springer},\ \bibinfo {year}
  {2002})\BibitemShut {NoStop}%
\bibitem [{\citenamefont {Rolland}(1993)}]{RollandDiss}%
  \BibitemOpen
  \bibfield  {author} {\bibinfo {author} {\bibfnamefont {A.}~\bibnamefont
  {Rolland}},\ }\emph {\bibinfo {title} {Contribution \`a l'analyse physique du
  transistor couches minces \`a base de silicium amorphe hydrog\'en\'e}},\
  \href@noop {} {Ph.D. thesis},\ \bibinfo  {school} {University of Rennes}
  (\bibinfo {year} {1993})\BibitemShut {NoStop}%
\bibitem [{\citenamefont {Rolland}\ \emph {et~al.}(1993)\citenamefont
  {Rolland}, \citenamefont {Richard}, \citenamefont {Kleider},\ and\
  \citenamefont {Mencaraglia}}]{Rolland1993}%
  \BibitemOpen
  \bibfield  {author} {\bibinfo {author} {\bibfnamefont {A.}~\bibnamefont
  {Rolland}}, \bibinfo {author} {\bibfnamefont {J.}~\bibnamefont {Richard}},
  \bibinfo {author} {\bibfnamefont {J.~P.}\ \bibnamefont {Kleider}}, \ and\
  \bibinfo {author} {\bibfnamefont {D.}~\bibnamefont {Mencaraglia}},\
  }\href@noop {} {\bibfield  {journal} {\bibinfo  {journal} {J. Electrochem.
  Soc.}\ }\textbf {\bibinfo {volume} {140}},\ \bibinfo {pages} {3679} (\bibinfo
  {year} {1993})}\BibitemShut {NoStop}%
\bibitem [{\citenamefont {Kim}\ \emph {et~al.}(2013)\citenamefont {Kim},
  \citenamefont {Castro-carranza}, \citenamefont {Estrada}, \citenamefont
  {Cerdeira}, \citenamefont {Bonnassieux}, \citenamefont {Horowitz},\ and\
  \citenamefont {I\~{n}iguez}}]{Kim2013}%
  \BibitemOpen
  \bibfield  {author} {\bibinfo {author} {\bibfnamefont {C.~H.}\ \bibnamefont
  {Kim}}, \bibinfo {author} {\bibfnamefont {A.}~\bibnamefont
  {Castro-carranza}}, \bibinfo {author} {\bibfnamefont {M.}~\bibnamefont
  {Estrada}}, \bibinfo {author} {\bibfnamefont {A.}~\bibnamefont {Cerdeira}},
  \bibinfo {author} {\bibfnamefont {Y.}~\bibnamefont {Bonnassieux}}, \bibinfo
  {author} {\bibfnamefont {G.}~\bibnamefont {Horowitz}}, \ and\ \bibinfo
  {author} {\bibfnamefont {B.}~\bibnamefont {I\~{n}iguez}},\ }\href@noop {}
  {\bibfield  {journal} {\bibinfo  {journal} {IEEE Transactions on Electron
  Devices}\ }\textbf {\bibinfo {volume} {60}},\ \bibinfo {pages} {1136}
  (\bibinfo {year} {2013})}\BibitemShut {NoStop}%
\bibitem [{\citenamefont {Xu}, \citenamefont {Balestra},\ and\ \citenamefont
  {Ghibaudo}(2011)}]{Xu2011}%
  \BibitemOpen
  \bibfield  {author} {\bibinfo {author} {\bibfnamefont {Y.}~\bibnamefont
  {Xu}}, \bibinfo {author} {\bibfnamefont {F.}~\bibnamefont {Balestra}}, \ and\
  \bibinfo {author} {\bibfnamefont {G.}~\bibnamefont {Ghibaudo}},\ }\href
  {\doibase 10.1063/1.3599485} {\bibfield  {journal} {\bibinfo  {journal}
  {Applied Physics Letters}\ }\textbf {\bibinfo {volume} {98}},\ \bibinfo {eid}
  {233302} (\bibinfo {year} {2011})}\BibitemShut {NoStop}%
\bibitem [{\citenamefont {Fichtner1979}, \citenamefont {Fichtner}(1979)}]{Fichtner1979}%
  \BibitemOpen
  \bibfield  {author} {\bibinfo {author} {\bibfnamefont {W.}~\bibnamefont
  {Fichtner}}, and \bibinfo {author} {\bibfnamefont {H. W.}~\bibnamefont {P\"{o}tzl}}, \ } {\bibfield  {journal} {\bibinfo  {journal}
  {Int. J. Electron.}\ }\textbf {\bibinfo {volume} {46}},\ \bibinfo
  {pages} {33} (\bibinfo {year} {1979})}\BibitemShut {NoStop}%
\bibitem [{\citenamefont {Oberhoff}\ \emph {et~al.}(2007)\citenamefont
  {Oberhoff}, \citenamefont {Pernstich}, \citenamefont {Member}, \citenamefont
  {Gundlach},\ and\ \citenamefont {Batlogg}}]{Oberhoff2007}%
  \BibitemOpen
  \bibfield  {author} {\bibinfo {author} {\bibfnamefont {D.}~\bibnamefont
  {Oberhoff}}, \bibinfo {author} {\bibfnamefont {K.~P.}\ \bibnamefont
  {Pernstich}}, \bibinfo {author} {\bibfnamefont {S.}~\bibnamefont {Member}},
  \bibinfo {author} {\bibfnamefont {D.~J.}\ \bibnamefont {Gundlach}}, \ and\
  \bibinfo {author} {\bibfnamefont {B.}~\bibnamefont {Batlogg}},\ }\href@noop
  {} {\bibfield  {journal} {\bibinfo  {journal} {IEEE Transactions on Electron
  Devices}\ }\textbf {\bibinfo {volume} {54}},\ \bibinfo {pages} {17} (\bibinfo
  {year} {2007})}\BibitemShut {NoStop}%
\bibitem [{\citenamefont {Scheinert}\ \emph {et~al.}(2007)\citenamefont
  {Scheinert}, \citenamefont {Pernstich}, \citenamefont {Batlogg},\ and\
  \citenamefont {Paasch}}]{Scheinert2007}%
  \BibitemOpen
  \bibfield  {author} {\bibinfo {author} {\bibfnamefont {S.}~\bibnamefont
  {Scheinert}}, \bibinfo {author} {\bibfnamefont {K.~P.}\ \bibnamefont
  {Pernstich}}, \bibinfo {author} {\bibfnamefont {B.}~\bibnamefont {Batlogg}},
  \ and\ \bibinfo {author} {\bibfnamefont {G.}~\bibnamefont {Paasch}},\ }\href
  {\doibase 10.1063/1.2803742} {\bibfield  {journal} {\bibinfo  {journal}
  {Journal of Applied Physics}\ }\textbf {\bibinfo {volume} {102}},\ \bibinfo
  {pages} {104503} (\bibinfo {year} {2007})}\BibitemShut {NoStop}%
\bibitem [{\citenamefont {Kim}, \citenamefont {Bonnassieux},\ and\
  \citenamefont {Horowitz}(2013)}]{Kim2013a}%
  \BibitemOpen
  \bibfield  {author} {\bibinfo {author} {\bibfnamefont {C.}~\bibnamefont
  {Kim}}, \bibinfo {author} {\bibfnamefont {Y.}~\bibnamefont {Bonnassieux}}, \
  and\ \bibinfo {author} {\bibfnamefont {G.}~\bibnamefont {Horowitz}},\ }\href
  {http://ieeexplore.ieee.org/xpls/abs\_all.jsp?arnumber=6361466} {\bibfield
  {journal} {\bibinfo  {journal} {IEEE Electron Device Letters}\ }\textbf
  {\bibinfo {volume} {60}},\ \bibinfo {pages} {280} (\bibinfo {year}
  {2013})}\BibitemShut {NoStop}%
\bibitem [{\citenamefont {Selberherr}(1984)}]{Selberherr1984}%
  \BibitemOpen
  \bibfield  {author} {\bibinfo {author} {\bibfnamefont {S.}~\bibnamefont
  {Selberherr}},\ }\href {http://books.google.ch/books?id=EE4HlRZTYi4C} {\emph
  {\bibinfo {title} {Analysis and Simulation of Semiconductor Devices}}}\
  (\bibinfo  {publisher} {Springer},\ \bibinfo {year} {1984})\BibitemShut
  {NoStop}%
\bibitem [{\citenamefont {Kalb}\ \emph {et~al.}(2010)\citenamefont {Kalb},
  \citenamefont {Haas}, \citenamefont {Krellner}, \citenamefont {Mathis},\ and\
  \citenamefont {Batlogg}}]{Kalb2010b}%
  \BibitemOpen
  \bibfield  {author} {\bibinfo {author} {\bibfnamefont {W.~L.}\ \bibnamefont
  {Kalb}}, \bibinfo {author} {\bibfnamefont {S.}~\bibnamefont {Haas}}, \bibinfo
  {author} {\bibfnamefont {C.}~\bibnamefont {Krellner}}, \bibinfo {author}
  {\bibfnamefont {T.}~\bibnamefont {Mathis}}, \ and\ \bibinfo {author}
  {\bibfnamefont {B.}~\bibnamefont {Batlogg}},\ }\href {\doibase
  10.1103/PhysRevB.81.155315} {\bibfield  {journal} {\bibinfo  {journal} {Phys.
  Rev. B}\ }\textbf {\bibinfo {volume} {81}},\ \bibinfo {pages} {155315}
  (\bibinfo {year} {2010})}\BibitemShut {NoStop}%
\bibitem [{\citenamefont {Jurchescu}, \citenamefont {Meetsma},\ and\
  \citenamefont {Palstra}(2006)}]{Jurchescu2006}%
  \BibitemOpen
  \bibfield  {author} {\bibinfo {author} {\bibfnamefont {O.~D.}\ \bibnamefont
  {Jurchescu}}, \bibinfo {author} {\bibfnamefont {A.}~\bibnamefont {Meetsma}},
  \ and\ \bibinfo {author} {\bibfnamefont {T.~T.~M.}\ \bibnamefont {Palstra}},\
  }\href {\doibase 10.1107/S0108768106003053} {\bibfield  {journal} {\bibinfo
  {journal} {Acta Crystallographica Section B}\ }\textbf {\bibinfo {volume}
  {62}},\ \bibinfo {pages} {330} (\bibinfo {year} {2006})}\BibitemShut
  {NoStop}%
\bibitem [{\citenamefont {Zhuo}\ \emph {et~al.}(2012)\citenamefont {Zhuo},
  \citenamefont {Sannomiya}, \citenamefont {Goto}, \citenamefont {Yamada},
  \citenamefont {Ohmi}, \citenamefont {Kakiuchi},\ and\ \citenamefont
  {Yasutake}}]{Zhuo2012}%
  \BibitemOpen
  \bibfield  {author} {\bibinfo {author} {\bibfnamefont {Z.}~\bibnamefont
  {Zhuo}}, \bibinfo {author} {\bibfnamefont {Y.}~\bibnamefont {Sannomiya}},
  \bibinfo {author} {\bibfnamefont {K.}~\bibnamefont {Goto}}, \bibinfo {author}
  {\bibfnamefont {T.}~\bibnamefont {Yamada}}, \bibinfo {author} {\bibfnamefont
  {H.}~\bibnamefont {Ohmi}}, \bibinfo {author} {\bibfnamefont {H.}~\bibnamefont
  {Kakiuchi}}, \ and\ \bibinfo {author} {\bibfnamefont {K.}~\bibnamefont
  {Yasutake}},\ }\href {\doibase 10.1016/j.cap.2012.04.015} {\bibfield
  {journal} {\bibinfo  {journal} {Current Applied Physics}\ }\textbf {\bibinfo
  {volume} {12}},\ \bibinfo {pages} {S57} (\bibinfo {year} {2012})}\BibitemShut
  {NoStop}%
\end{thebibliography}
\end{document}